\documentclass[3p,times,twocolumn]{elsarticle}

\usepackage{ecrc}

\volume{00}

\firstpage{1}

\journalname{Nuclear Physics B Proceedings Supplement}

\runauth{K. Hattori and K. Itakura}

\jid{nppp}

\jnltitlelogo{Nuclear Physics B Proceedings Supplement}

\biboptions{sort, square}

\usepackage[figuresright]{rotating}
\usepackage{latexsym}
\usepackage{amsfonts}
\usepackage{amssymb}
\usepackage{amsmath}
\usepackage{amsthm}
\usepackage{mathrsfs} 	
\usepackage{bm}
\usepackage{graphicx}
\usepackage{color}
\usepackage{ascmac}
\usepackage{time}
\usepackage{dcolumn}  
\usepackage{epsfig}

\newcommand{\bq}{{\bm{q}}}

\newcommand{\M}{ {\mathcal M} }

\usepackage[normalem]{ulem}  

\renewcommand\sout{\bgroup \color{red} \ULdepth=-.5ex \ULset}

\begin{document}

\begin{frontmatter}

\title{Photon and dilepton spectra from nonlinear QED effects 
in supercritical magnetic fields induced by heavy-ion collisions}

\author[rbrc,riken]{Koichi Hattori}
\ead{koichi.hattori@riken.jp}

\author[kek,soken]{Kazunori Itakura}
\ead{kazunori.itakura@kek.jp}

\address[rbrc]{RIKEN BNL Research Center, Brookhaven National Laboratory, Upton NY 11973, USA}
\address[riken]{Theoretical Research Division, Nishina Center, RIKEN, Wako, Saitama 351-0198, Japan}
\address[kek]{KEK Theory Center, IPNS, 
High Energy Accelerator Research Organization, 1-1, Oho, Ibaraki, 305-0801, Japan}
\address[soken]{Graduate University for Advanced Studies (SOKENDAI),
1-1 Oho, Tsukuba, Ibaraki 305-0801, Japan}

\begin{abstract}
We discuss properties of photons in extremely strong magnetic fields 
induced by the relativistic heavy-ion collisions. 
We investigate the vacuum birefringence, the real-photon decay, and the photon splitting 
which are all forbidden in the ordinary vacuum, but become possible in strong magnetic fields. 
These effects potentially give rise to anisotropies in photon and dilepton spectra. 
\end{abstract}


\end{frontmatter}



\section{Introduction}
\label{}

Electromagnetic probes are expected to be penetrating probes of the matter 
created in the ultrarelativistsic heavy-ion collisions. 
Especially, it is interesting to pursue the possibility 
if any aspect of the early-time dynamics can be probed. 
One of the ingredients recently bringing lots of excitements in the early-time dynamics is 
a strong magnetic field induced by the colliding nuclei \cite{SMS, KMW, Bestimates}. 
We would like to point out that the strong magnetic field gives rise to 
intriguing modifications of the photon properties which arise 
only in the presence of strong magnetic fields \cite{HI1,HI2,Ish,HI3}. 
Effects of the strong magnetic field, as well as the medium effects studied in, e.g., Refs.~\cite{LLWX,AM}, 
will be important for the optics in the heavy-ion collisions. 

\section{Charged-fermion spectrum and resummation in supercritical magnetic fields}

The important quantum effect of the magnetic field is 
the Landau-level discretization of the charged particle spectrum. 
The transverse energy level is discretized because of the periodic synchrotron motion, 
while the momentum longitudinal to the magnetic field is still continuous, 
so that the spectrum of charged fermion becomes anisotropic as $({\bm B} = (0,0,B))$
\begin{eqnarray}
E_n = \sqrt{ m^2 + p_z^2 + 2 n eB} \ \  \ \ (n \geq 0)
\label{eq:LL}
\, .
\end{eqnarray} 
This is particularly important to study the photon properties 
since the modification of the photon refractive index 
is entirely due to the quantum fluctuations of charged fermions in the Dirac sea 
responding to the propagating electromagnetic fields of photons. 

\begin{figure}[t]
     \begin{center}
              \includegraphics[width=0.7\hsize]{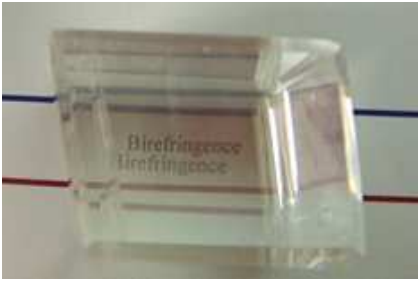}
     \end{center}
\vspace{-0.5cm}
\caption{
Doubled image through a birefringent substance ``calcite''. 
Rays split at the refraction on the surface of calcite 
because of a polarization dependence of refractive index. 
}
\label{fig:calcite}
\end{figure}

A consequence of the anisotropic spectrum can be seen 
in the {\it birefringence} shown in Fig.~\ref{fig:calcite}. 
Because of the anisotropic spectrum, the response of the charged particles 
depends on the polarization modes, i.e., the direction of the electric field, 
and thus the each polarization mode has a refractive index different from the other. 
As discussed below, an analogous effect called the {\it vacuum birefringence} 
will be found in the presence of the magnetic field even without any medium. 

To properly include effects of  the Landau-level discretization, 
one needs to carry out the resummation with respect to 
the number of external-field insertions as shown in Fig.~\ref{fig:resummation}. 
By using the resummed propagator (for explicit forms, see, e.g., \cite{HI1,GMS}), 
the anisotropic spectrum can be taken into account in the computation of the optical properties. 

The useful tool for the resummation is the proper-time method \cite{Sch} 
which has been used to show the phenomena seen in the strong external fields 
such as the Schwinger pair creation in strong electric fields. 
In the context of heavy-ion physics, the pair creation in the Coulomb field of high-Z atoms 
were studied in early days (see Ref.~\cite{LM} and references therein). 
Such quantum phenomena will become sizable effects 
when the strengths of the external electromagnetic fields exceed 
the critical field strength $B_c = m^2/e$ 
for the mass of the vacuum fluctuation and the coupling constant.


\section{Vacuum birefringence and real-photon decay}

\begin{figure}[t]
     \begin{center}
              \includegraphics[width=0.9\hsize]{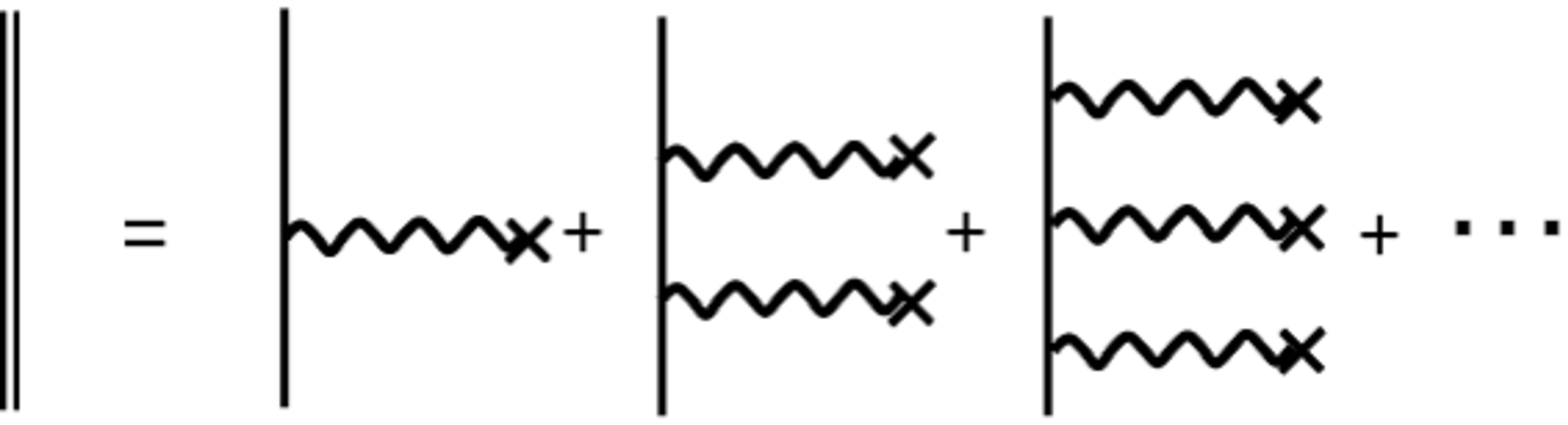}
     \end{center}
\vspace{-0.7cm}     
\caption{Resummation with respect to external-field insertions.}
\label{fig:resummation}
     \begin{center}
              \includegraphics[width=\hsize]{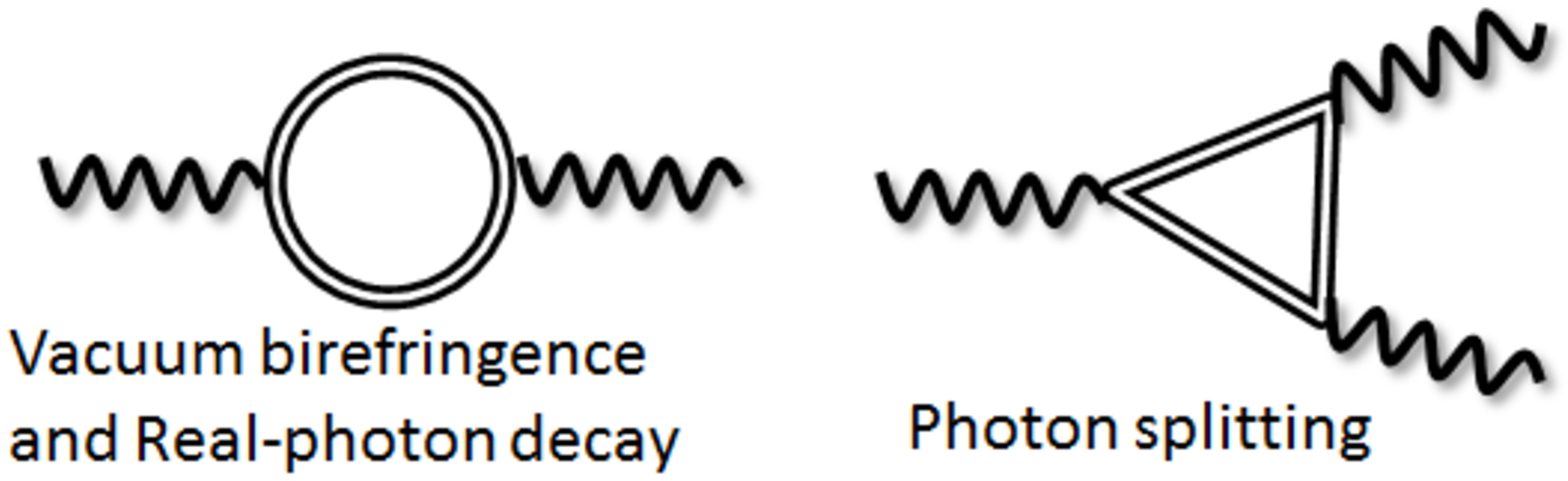}
     \end{center}
\vspace{-0.7cm}     
\caption{Polarization tensor and triangle diagram with the resummed propagator.}
\label{fig:diagrams}
\end{figure}

Refractive indices can be obtained 
from the vacuum polarization diagram in Fig.~\ref{fig:diagrams}. 
In the ordinary vacuum without external field, this quantum fluctuation does not give rise to 
any modification of the refractive index, because of the Lorentz and gauge symmetries. 
However, the external magnetic field breaks the Lorentz symmetry, 
and then the refractive indices get non-trivial modifications 
such as the polarization dependence discussed above 
due to the anisotropic fermion spectrum (\ref{eq:LL}) from the Landau-level discretization.


\subsection{Analytic computation of the polarization tensor}

\begin{figure}
     \begin{center}
              \includegraphics[width=0.9\hsize]{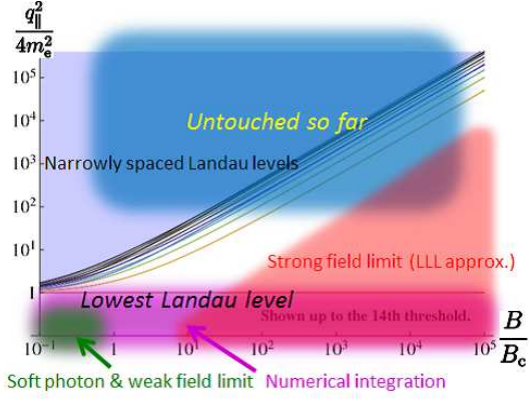}
     \end{center}
\vspace{-0.5cm}          
\caption{Summary of the relevant scales in the problem, 
the field strength and the photon momentum.}
\label{fig:scales}
\end{figure}

The general form of the polarization tensor 
is complicated due to the resummation. 
Nevertheless, one can get simple results in some particular limits. 
Figure~\ref{fig:scales} shows a summary of  the relevant scales in the problem 
which are specified by the strength of the magnetic field and the photon momentum. 
For example, one can get a useful approximation in the strong field limit by 
including a contribution of only the lowest Landau level (LLL) \cite{Fuk, HI1, HI2}. 
However, the validity of the approximation also depends on the other scale, i.e., the photon momentum, 
and the LLL approximation will not suffice when the momentum scale becomes large. 
Actually, the regime of the large momentum and the strong field is relevant for 
the ultrarelativistic heavy-ion collisions. 
As far as we know, there has not been the general result 
which is valid in such a regime.

The formal expression of the resummed polarization tensor has 
a gauge-invariant form 
\begin{eqnarray}
\Pi^{\mu\nu} (q^2) = - ( \chi_0 P^{\mu\nu} +  \chi_1 P_\parallel^{\mu\nu} 
+  \chi_2 P_\perp^{\mu\nu} )
\end{eqnarray}
where the transverse projection operators 
$P^{\mu\nu} = q^2 g^{\mu\nu} - q^\mu q^\nu$, 
$P_\parallel^{\mu\nu} = q_\parallel^2 g^{\mu\nu}_\parallel - q_\parallel^\mu q_\parallel^\nu$, 
and $P_\perp^{\mu\nu} = q_\perp^2 g^{\mu\nu}_\perp - q_\perp^\mu q_\perp^\nu$ 
are defined by the metrics in the longitudinal and transverse subspaces 
$g_\parallel^{\mu\nu}= {\rm diag} (1,0,0,-1)$ and $g_\parallel^{\mu\nu}= {\rm diag} (0,-1,-1,0)$, 
and the longitudinal and transverse momenta $q_\parallel^\mu = g_\parallel^{\mu\nu}q_\mu$ 
and $q_\perp^\mu = g_\perp^{\mu\nu}q_\mu$. 
Here, the magnetic field is applied in the third direction.

We could perform analytic computation of the coefficient functions $\chi_{0,1,2}$ 
by using relations among special functions \cite{HI1}. 
The analytic results are expressed by the wave functions of charged particles, 
namely the associated Laguerre polynomials, which naturally arise in the calculation, 
and by the summation with respect to the contributions of the Landau levels. 
The exact expressions of the resummed polarization tensor were missing in the last few decades 
(c.f. earlier attempts in Ref.~\cite{KRDG}). 
Our general result covers the whole parameter region in Fig.\,\ref{fig:scales}, 
and will be useful to study a wide variety of systems which contain quite different scales, 
e.g., heavy-ion collisions, neutron star/magnetars, early universe, high-intensity laser field, etc.

\subsection{Refractive index in the LLL}

By using the polarization tensor, we show the refractive index in the strong magnetic field \cite{HI2}. 
To investigate the basic features, let us focus on the lowest Landau level 
which can be simply generalized by including the contribution of relevant Landau levels 
contained in the analytic result of the polarization tensor discussed above. 

\begin{figure}
     \begin{center}
              \includegraphics[width=0.5\hsize]{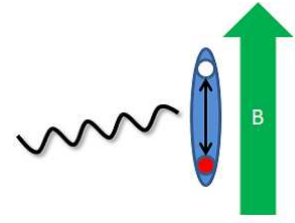}
     \end{center}
\vspace{-0.7cm}         
\caption{1+1 dimensional fluctuation in the lowest Landau level.}
\label{fig:LLL}
\end{figure}

In the LLL approximation \cite{Fuk, HI1, HI2}, 
two of the three coefficient functions are vanishing $\chi_0=\chi_2=0$. 
This can be understood from the dimensional reduction in the LLL \cite{GMS} 
where the charged fermions can fluctuate 
only in the direction of the external magnetic field (see Fig.~\ref{fig:LLL}). 
The nonvanishing component $\chi_1$ has a simple form 
\begin{eqnarray}
&& \hspace{-1cm}
\chi_{\rm LLL} = \frac{e^2}{\pi} \cdot \frac{eB}{2\pi} e^{-\frac{\vert \bq_\perp \vert^2}{eB}} 
q_\parallel^{-2} \{ I(q_\parallel^2) - 1 \}
,
\\
&& \hspace{-1cm}
I(q_\parallel^2) =
\frac{4m^2}{ \sqrt{q_\parallel^2(4m^2 - q_\parallel^2) } }
\arctan  \sqrt{ \frac{q_\parallel^2 } {(4m^2 -q_\parallel^2) } } 
,
\end{eqnarray}
where the above form of $I(q_\parallel^2)$ is valid when $0 \leq q_\parallel^2 < 4m^2$ 
and can be analytically continued to the other regions $q_\parallel^2 < 0$ and $4m^2 \leq q_\parallel^2$. 
A remarkable feature is that the $\chi_1$ acquires an imaginary part 
from the arctangent when the momentum goes beyond 
the invariant mass of the fermion and antifermion pair $4m^2 \leq q_\parallel^2$. 
This indicates decay of a real photon in the presence of external magnetic fields. 

The dilelectric constant is obtained from the solution of the Maxwell equation 
or equivalently from the pole position of the resummed photon propagator 
with insertions of the ring diagrams \cite{HI1} as 
\begin{eqnarray}
\epsilon = n^2 = \frac{1+\chi_{\rm LLL}}{1+\chi_{\rm LLL} \cos^2\theta} 
\end{eqnarray}
When the dielectric constant has the imaginary part, 
$\epsilon = \epsilon_{\rm real} + i \, \epsilon_{\rm imag}$, 
the refractive index also has both real and imaginary parts $n = n_{\rm real} + i \, n_{\rm imag}$ 
which are related as 
\begin{eqnarray}
\label{eq:ref_real}
&& n_{\rm real} \ = 
\ \frac{1}{\sqrt{2}} \sqrt{  | \epsilon | + \epsilon_{\rm real}  }\, ,  
\\
&& n_{\rm imag} \ = 
\ \frac{1}{\sqrt{2}} \sqrt{  | \epsilon | - \epsilon_{\rm real}  }  
\, ,
\label{eq:ref_imag}
\end{eqnarray}
with the absolute value $| \epsilon |=\sqrt{ \epsilon_{\rm real} ^2 + \epsilon_{\rm imag} ^2 }$.

When photons are propagating in the supercritical magnetic fields, 
the real part of the refractive index can be, e.g., $\sim 1.4 $ for $B/B_c = 500$ 
which is much larger than the refractive index of air (1.0003) 
and is comparable to that of water (1.333). 

\begin{figure}
     \begin{center}
              \includegraphics[width=0.75\hsize]{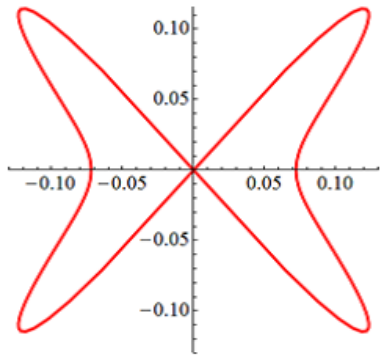}
     \end{center}
\vspace{-0.7cm}        
\caption{The imaginary part of the refractive index at the field strength $B/B_c = 500$ 
and the photon energy $\omega^2/ (4m^2) = 3$.}
\label{fig:imag}
\end{figure}

Figure~\ref{fig:imag} shows the imaginary part of the refractive index. 
The magnetic field is applied in the vertical direction, and an angle measured from 
the vertical axis corresponds to the angle between the direction of the magnetic field 
and the momentum of a photon. A distance from the origin to the red curve indicates 
the magnitude of the imaginary part. The angle dependence of the red curve indicates 
an anisotropy of the imaginary part. In this figure, the decay rate is large 
when the photons are propagating with angles $\sim \pm \pi/4 , \, \pm 3\pi/4$, 
which will potentially give rise the anisotropic spectrum of 
the direct photons in the ultrarelativistic heavy-ion collisions.

\section{Photon splitting}

We discuss another intriguing phenomenon called the photon splitting 
shown in Fig.~\ref{fig:diagrams}. 
In the ordinary vacuum, 
the splitting of a photon into two photons is prohibited, 
because Furry's theorem tells us that the two independent diagrams 
with clockwise and counterclockwise charge flows cancel 
for odd orders of the vector current correlators. 
However, with the help of external magnetic fields, 
the splitting process becomes possible because the diagrams could have in total 
an even number of external legs which are provided by an odd number of the external magnetic field 
and the three dynamical photon lines. 
Although these diagrams are higher order in the naive order counting of the coupling constant, 
the strong magnetic field compensates the suppression.

In the strong magnetic field limit, the dominant contribution 
would come from the triangle diagram composed of the three lowest Landau levels. 
However, we showed that this is not the case. 
As shown in Fig.~\ref{fig:LLL}, these 1+1 dimensional fluctuations can couple only to 
the polarization mode oscillating in the direction of the magnetic field. 
However, this splitting ($\parallel \to \parallel + \parallel$) 
is not allowed due to the polarization selection rule \cite{Adl}
\footnote{Note the notations of $\parallel$ and $\perp$ in the literature.}.


Therefore, the leading contribution in the strong magnetic field limit 
comes from the configuration with the two lowest Landau levels and 
one higher Landau level [denoted with a short red line in Eq.~(\ref{eq:splitting})]. 
The energy level of the latter depends on 
the magnitude of magnetic field [$n \geq 1$ in Eq.~(\ref{eq:LL})] 
which can be seen as a large effective ''mass'' $M^2 = 2 eB$. 
In the low energy regime where the external photon momenta are much smaller than $M^2$, 
the far-off-shell ``heavy'' fermion propagates only in a short distance, 
shrinking into an effective coupling like the four-Fermi interaction for the weak bosons. 
Then, the order of the photon-splitting amplitude reads
\setbox1\hbox to
2.cm{\resizebox*{2cm}{!}{\includegraphics{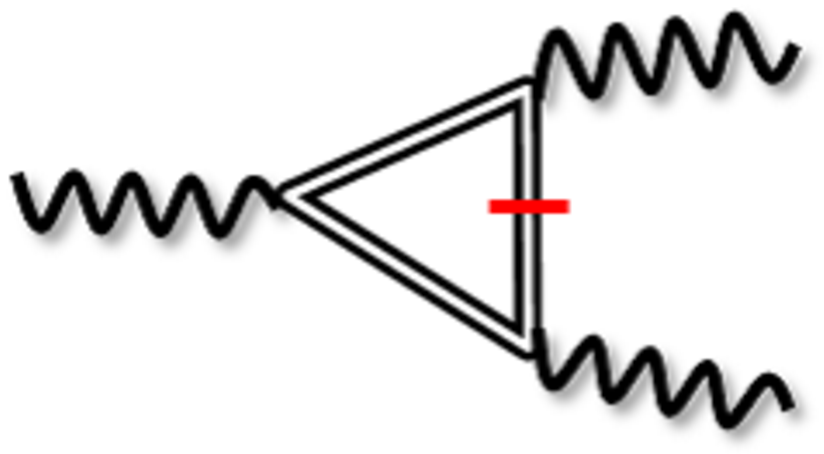}}}
\setbox2\hbox to
2.cm{\resizebox*{2cm}{!}{\includegraphics{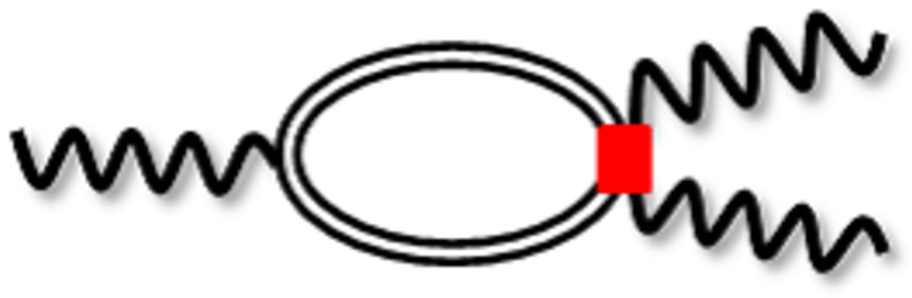}}}
\begin{equation}
\M =
\raise - 4.5mm\box1
\sim
\raise -4.5mm\box2
\sim e^3 \frac{{\sqrt{2eB\;}}^2}{M^2}
\label{eq:splitting}
\, .
\end{equation}
A red vertex in the second diagram shows the effective coupling $e^2/M^2$, 
and a factor of ${\sqrt{2eB\;}}^2$ comes from the density of state 
in the transverse subspace. 
The order of $\M$ is $ {\mathcal O}\left( \, e^3 (eB)^0 \, \right)$, 
indicating that the photon-splitting rate in the strong field limit is 
independent of the magnitude of magnetic field. 
This may clarify earlier observations in the strong field limit \cite{BMS, CKM}. 
Details of our investigation will appear elsewhere \cite{HI3}.



\section{Concluding remarks}

We discussed the photon properties in the strong magnetic fields induced 
by the ultrarelativistic heavy-ion collisions. 
We would like to note that these phenomena have been discussed also 
in astrophysics \cite{HL, Bar} and intense-laser physics \cite{laser}. 
Heavy-ion physics could be the first to observe such phenomena 
and get an impact on the interdisciplinary study.



\vspace{0.5cm}

{\bf Acknowledgements} ---
The research of K.H. is supported by JSPS Grants-in-Aid No.~25287066.


\begin{thebibliography}{00}

\bibitem{SMS} 
S. Schramm, B. Muller, and A. J. Schramm, Mod. Phys. Lett. A {\bf 7}, 973 (1992); 
Phys. Lett. A {\bf 164}, 28 (1992). 

\bibitem{KMW} D. E. Kharzeev, L. D. McLerran, and H. J. Warringa, 
Nucl. Phys. {\bf A803}, 227 (2008). 

\bibitem{Bestimates} V. Skokov, A. Y. Illarionov, and V. Toneev, 
Int. J. Mod. Phys. A {\bf 24}, 5925 (2009); 
W.~T. Deng and X.~G. Huang, Phys. Rev. C {\bf 85}, 044907 (2012).


\bibitem{HI1} K.~Hattori and K.~Itakura,
  Annals Phys.\  {\bf 330}, 23 (2013)
  [arXiv:1209.2663 [hep-ph]]. 

\bibitem{HI2}  K.~Hattori and K.~Itakura,
  Annals Phys.\  {\bf 334}, 58 (2013) 
  [arXiv:1212.1897 [hep-ph]].
  
\bibitem{Ish}   K.~I.~Ishikawa, D.~Kimura, K.~Shigaki and A.~Tsuji,
  Int.\ J.\ Mod.\ Phys.\ A {\bf 28}, 1350100 (2013)
  [arXiv:1304.3655 [hep-ph]].
 
\bibitem{HI3} K.~Hattori and K.~Itakura, In preparation.
 

\bibitem{LLWX}   J.~Liu, M.~j.~Luo, Q.~Wang and H.~j.~Xu,
  Phys.\ Rev.\ D {\bf 84}, 125027 (2011)
  [arXiv:1109.4083 [hep-ph]].

\bibitem{AM}   A.~Monnai,
  arXiv:1408.1410 [nucl-th]. 
  
  
\bibitem{GMS} V. P. Gusynin, V. A. Miransky and I. A. Shovkovy, 
Nucl. Phys. {\bf B462}, 249 (1996). 

\bibitem{Sch}   J.~S.~Schwinger,
  Phys.\ Rev.\  {\bf 82}, 664 (1951).

\bibitem{LM}   L.~McLerran,
  Int.\ J.\ Mod.\ Phys.\ E {\bf 16}, 805 (2007).

  
\bibitem{Fuk}   K.~Fukushima,
  Phys.\ Rev.\ D {\bf 83}, 111501 (2011)
  [arXiv:1103.4430 [hep-ph]].  

  
\bibitem{KRDG}   F.~Karbstein, L.~Roessler, B.~Dobrich and H.~Gies,
  Int.\ J.\ Mod.\ Phys.\ Conf.\ Ser.\  {\bf 14}, 403 (2012)
  [arXiv:1111.5984 [hep-ph]].

  
  
\bibitem{Adl}   S.~L.~Adler, J.~N.~Bahcall, C.~G.~Callan and M.~N.~Rosenbluth,
  Phys.\ Rev.\ Lett.\  {\bf 25}, 1061 (1970);
    S.~L.~Adler,
  Annals Phys.\  {\bf 67}, 599 (1971);
S.~L.~Adler and C.~Schubert,
   Phys.\ Rev.\ Lett.\  {\bf 77}, 1695 (1996)
   [hep-th/9605035].
  
  \bibitem{BMS}   V.~N.~Baier, A.~I.~Milshtein and R.~Z.~Shaisultanov,
  Phys.\ Rev.\ Lett.\  {\bf 77}, 1691 (1996)
  
  \bibitem{CKM}   M.~V.~Chistyakov, A.~V.~Kuznetsov and N.~V.~Mikheev,
  Phys.\ Lett.\ B {\bf 434}, 67 (1998)
  [hep-ph/9804444].
  
  
  \bibitem{HL}   A.~K.~Harding and D.~Lai,
  Rept.\ Prog.\ Phys.\  {\bf 69}, 2631 (2006)
  [astro-ph/0606674].
  
  \bibitem{Bar}   M.~G.~Baring,
  AIP Conf.\ Proc.\  {\bf 1051}, 53 (2008)
  [arXiv:0804.0832 [astro-ph]].

\bibitem{laser} ``Topical issue on Fundamental physics and ultra-high laser fields'', Eur. Phys. J. D {\bf 55}  (2009).

\end{thebibliography}
\end{document}